\documentclass[10pt,onecolumn]{iontran}

\usepackage[usenames]{xcolor}

\usepackage{amsmath,amsfonts,amscd,amssymb,latexsym}
\usepackage[active]{srcltx}  
\usepackage{graphicx}       
\usepackage{fancyhdr,url}
\usepackage{dblfloatfix}  
\usepackage{changepage}
\usepackage{lipsum}
\usepackage{multirow}
\usepackage{booktabs}
\usepackage{afterpage}
\newcommand{\thickhline}{
    \noalign {\ifnum 0=`}\fi \hrule height 1pt
    \futurelet \reserved@a \@xhline
}
\usepackage{subfigure,wrapfig,graphicx,booktabs,fancyhdr,amsmath, amsfonts}
\usepackage{bm, amssymb,amsmath,amsthm}
\usepackage{makecell}
\usepackage{enumitem}
\usepackage{epstopdf}
\usepackage[letterpaper, left=.75in,right=.75in,top=1in,bottom=1in,]{geometry}

\usepackage{epstopdf}

\newcommand{\barr}{\begin{array}}
\newcommand{\earr}{\end{array}}
\newcommand{\ba}{\begin{array}}
\newcommand{\ea}{\end{array}}
\newcommand{\beq}{\begin{equation}}
\newcommand{\eeq}{\end{equation}}
\newcommand{\be}{\begin{equation}}
\newcommand{\ee}{\end{equation}}
\newcommand{\beqa}{\begin{eqnarray}}
\newcommand{\eeqa}{\end{eqnarray}}
\newcommand{\bea}{\begin{eqnarray}}
\newcommand{\eea}{\end{eqnarray}}
\newcommand{\bc}{\begin{center}}
\newcommand{\ec}{\end{center}}
\newcommand{\bi}{\begin{itemize}}
\newcommand{\ei}{\end{itemize}}
\newcommand{\bd}{\begin{description}}
\newcommand{\ed}{\end{description}}
\newcommand{\bn}{\begin{enumerate}}
\newcommand{\en}{\end{enumerate}}
\newcommand{\bq}{\begin{quote}}
\newcommand{\eq}{\end{quote}}
\newcommand{\bv}{\begin{verbatim}}
\newcommand{\ev}{\end{verbatim}}

\def\litem[#1]{\item[#1\hfill]}   

\input epsf

\def\captionwidth#1{%
  \dimen0=\columnwidth   \advance\dimen0 by-#1\relax
  \divide\dimen0 by2
  \advance\leftskip by\dimen0
  \advance\rightskip by\dimen0
}

\usepackage{amsthm}
\usepackage{cite}
\usepackage{etoolbox}
\makeatletter
\patchcmd{\@makecaption}
  {\scshape}
  {}
  {}
  {}
\makeatother
\newcommand{\norm}[1]{\left\lVert#1\right\rVert_2}

\newcommand{\vb}{\boldsymbol}

\newcommand{\mb}[1]{{\mathbf{#1}}}

\newcommand{\trp}[1]{{#1}^{\mathsf{T}}}

\graphicspath{./figs/}

\usepackage{amsthm}
\usepackage{multirow}
\usepackage{booktabs}
\usepackage{authblk}

\usepackage{algorithm}
\usepackage{algorithmic}

\begin{document}

\title{Integrity-Based Path Planning Strategy for Urban Autonomous Vehicular Navigation \\ Using GPS and Cellular Signals}

\author[1]{\normalsize Halim Lee}
\author[1]{\normalsize Jiwon Seo}
\author[2]{\normalsize Zaher M. Kassas}
\affil[1]{\textit{Yonsei University, Korea}}
\affil[2]{\textit{University of California, Irvine, USA} }

\maketitle
\pagestyle{plain}
\thispagestyle{fancy}
\pagestyle{plain}
\thispagestyle{fancy} 
\rfoot{\footnotesize \bf Preprint of the 2020 ION GNSS+ Conference\\St. Louis, Missouri, September 21--25, 2020}

\lfoot{\footnotesize \bf Copyright \copyright~2020 by H. Lee, J. Seo, \\ and Z. Kassas}

\setlength{\parskip}{10pt plus0pt minus2pt}
\setlength{\parindent}{0mm}
\section*{BIOGRAPHIES}
\vspace{-0.30in}
\noindent \newline Halim Lee is an M.S./Ph.D. student in the School of Integrated Technology, Yonsei University, Korea. She received the B.S. degree in Integrated Technology from Yonsei University. She is currently a visiting graduate student at the Autonomous Systems Perception, Intelligent, and Navigation (ASPIN) Laboratory at the University of California, Irvine. Her research interests include motion planning, integrity monitoring, and opportunistic navigation.

\vspace{-0.15in}
\noindent \newline Jiwon Seo received the B.S. degree in mechanical engineering (division of aerospace engineering) in 2002 from the Korea Advanced Institute of Science and Technology (KAIST), Daejeon, Korea, and the M.S. degree in aeronautics and astronautics in 2004, the M.S. degree in electrical engineering in 2008, and the Ph.D. degree in aeronautics and astronautics in 2010 from Stanford University, Stanford, CA, USA. He is currently an Associate Professor with the School of Integrated Technology, Yonsei University, Korea. He is a member of the International Advisory Council of the Resilient Navigation and Timing Foundation, Alexandria, VA, USA, and a member of several advisory committees of the South Korean government.

\vspace{-0.15in}
\noindent \newline Zaher (Zak) M. Kassas is is an associate professor at the University of California, Irvine and director of the ASPIN Laboratory. He received a B.E. in Electrical Engineering from the Lebanese American University, an M.S. in Electrical and Computer Engineering from The Ohio State University, and an M.S.E. in Aerospace Engineering and a Ph.D. in Electrical and Computer Engineering from The University of Texas at Austin. In 2018, he received the National Science Foundation (NSF) Faculty Early Career Development Program (CAREER) award, and in 2019, he received the Office of Naval Research (ONR) Young Investigator Program (YIP) award. He is a recipient of 2018 IEEE Walter Fried Award, 2018 Institute of Navigation (ION) Samuel Burka Award, and 2019 ION Col. Thomas Thurlow Award. He is an Associate Editor for the IEEE Transactions on Aerospace and Electronic Systems and the IEEE Transactions on Intelligent Transportation Systems. His research interests include cyber-physical systems, estimation theory, navigation systems, autonomous vehicles, and intelligent transportation systems.

\section*{ABSTRACT}
\vspace{-0.15in}
An integrity-based path planning strategy for autonomous ground vehicle (AGV) navigation in urban environments is developed. The vehicle is assumed to navigate by utilizing cellular long-term evolution (LTE) signals in addition to Global Positioning System (GPS) signals. Given a desired destination, an optimal path is calculated, which minimizes a cost function that considers both the horizontal protection level (HPL) and travel distance. The constraints are that (i) the ratio of nodes with faulty signals to the total nodes be lower than a maximum allowable ratio and (ii) the HPLs along each candidate path  be lower than the horizontal alert limit (HAL). To predict the faults and HPL before the vehicle is driven, GPS and LTE pseudoranges along the candidate paths are generated utilizing a commercial ray-tracing software and three-dimensional (3D) terrain and building maps. Simulated pseudoranges inform the path planning algorithm about potential biases due to reflections from buildings in urban environments. Simulation results are presented showing that the optimal path produced by the proposed path planning strategy has the minimum average HPL among the candidate paths.
\vspace{-0.05in}
\section{INTRODUCTION}
To enhance the safety and convenience of driving, autonomous ground vehicles (AGVs) have been extensively researched by both academia and industry \cite{Kim14:Multi,CB17:Corporations,Rhee19:Low,Kassas17:Robust}. From a safety perspective, autonomous driving technology is expected to reduce collisions by drastically reducing human-driver errors and negligence \cite{Paden16:Survey}. Safety is highly dependent on the vehicle's navigation system. The vehicle's safety depends on the reliability of the position solution calculated by the vehicle's navigation system. 

The majority of AGVs developed so far rely on global navigation satellite systems (GNSS), such as the Global Positioning System (GPS), to determine their absolute coordinates \cite{Kassas19:Navigation2}. However, GNSS signals are susceptible to interference that can be caused by nature (e.g., ionospheric disturbances owing to solar activity \cite{Seo11:Availability,Seo14:Future, Lee17:Monitoring}) or humans (e.g., jamming \cite{Grant09:GPS,Park18:Dual,Kassas20:I_am} and spoofing \cite{Gunther14:Survey,Ioannides16:Known,Psiaki16:GNSS}). Especially in urban environments, the accuracy and availability of GNSS position solutions can be significantly degraded due to multipath and non-line-of-sight (NLOS) conditions caused by tall buildings. To compensate for the weakness of GNSS, navigation based on signals of opportunity (SOPs) such as AM/FM radio signals \cite{McEllroy06:Navigation,Fang09:Is_FM}, cellular signals\cite{Kassas17:I_Hear,Khalife16:NavigationII,Shamaei17:Exploiting,Kang19:TOA}, WiFi\cite{Yang15:WiFi-based, Zhuang15:Evaluation,Makki15:Survey}, and low earth orbit (LEO) satellite signals\cite{Lawrence17:Navigation,Landry19:Iridium,Kassas19:New_Age,Kassas20:Opportunity} have been actively studied. Among various SOPs, cellular long-term evolution (LTE) signals are desirable in urban environments due to their high signal strength and geometric diversity. Furthermore, there is no cost to receive LTE downlink signals for navigation purposes. Recent studies demonstrated  meter-level and submeter-level accuracy for cellular-signal-based ground vehicle \cite{Shamaei17:LTE,Shamaei19:AFramework} and unmanned aerial vehicle (UAV) \cite{Khalife18:Precise,Shamaei19:Submeter} navigation, respectively.

In order to ensure the safety of AGVs, it is necessary to continuously monitor the reliability of the navigation solution calculated by the vehicle's navigation system so that navigation integrity is guaranteed. The definition of integrity is ``the ability of the navigation system to provide timely warnings to the user when it is inadvisable to use the system for navigation \cite{Gebre09:GNSS}''. In aviation, three architectures are typically used to guarantee integrity: ground-based augmentation systems (GBAS), satellite-based augmentation systems (SBAS), and receiver autonomous integrity monitoring (RAIM). Among the three architectures, GBAS and SBAS rely on additional information from external sources (e.g., ground stations and satellites). On one hand, in open-sky environments, GBAS and SBAS can provide high navigation availability while guaranteeing  integrity, but their performances degrade in urban environments \cite{Zhu18:GNSS}. On the other hand, RAIM can guarantee the integrity by using redundant pseudorange measurements. RAIM can be used to detect faults with redundant pseudorange measurements and calculate the protection level, which is defined as a statistical error bound containing the position of the receiver with an acceptable level of certainty. Since it does not require additional information from external sources, RAIM is applicable even in harsh environments, such as urban areas \cite{Zhu18:GNSS}.

In this paper, integrity is considered in the context of path planning. Generally, the purpose of path planning is to find an optimal path from a start point to a target point, which can optimize a cost function generated in accordance with desired conditions (e.g., travel distance, traffic, and toll). Numerous path planning studies for autonomous vehicles have been conducted \cite{Nikolos03:Evolutionary,Chao17:Offline,Ragothaman19:Multipath}. In studies related to multi-objective path planning, path planning strategies that considered multiple objectives (e.g., obstacles, path length, and path smoothness) were proposed \cite{Alejandro17:Solving}. In studies related to simultaneous localization and mapping (SLAM), path planning strategies for minimizing the accumulated robot pose uncertainty \cite{Valencia18:Path} or minimizing the probability of becoming lost \cite{Valencia13:Planning} were suggested. In \cite{Maaref20:Optimal}, an integrity-constrained path planning strategy that considered both the travel distance and horizontal protection level (HPL) was proposed. This paper extends \cite{Maaref20:Optimal} in the following aspects:
\begin{list}{$\bullet$}{\leftmargin=1em \itemindent=0em}
  \item To increase the quality of the navigation solution (i.e., decrease the HPLs), cellular signals are used together with GPS signals, whereas only GPS signals were considered in \cite{Maaref20:Optimal}.
  \item Pseudorange bias caused by signal reflection by buildings is considered.
  \item Multiple faults in pseudorange measurements are considered in fault detection and HPL calculation. This is more realistic than a single fault assumption, which was used in \cite{Maaref20:Optimal}, since multiple faults can occur frequently in urban environments owing to signal reflections from buildings.
\end{list}

This paper is organized as follows. Section \ref{Section:Overview_of_Path_Planning_Method} gives an overview of the proposed path planning strategy. Section \ref{Section:GPS and LTE Measurements} presents GPS and LTE pseudorange prediction methods. Section \ref{Section:HPL Calculation} discusses fault prediction and HPL calculation methods. Section \ref{Section:Optimal Path Selection} proposes an optimal path planning method. Section \ref{Section:Simulation Results} simulation results. Section \ref{Section:Conclusions} concludes this paper.


\section{OVERVIEW OF PATH PLANNING STRATEGY} \label{Section:Overview_of_Path_Planning_Method}

\begin{figure}[h!]
\centering
\includegraphics[width=0.8\linewidth]{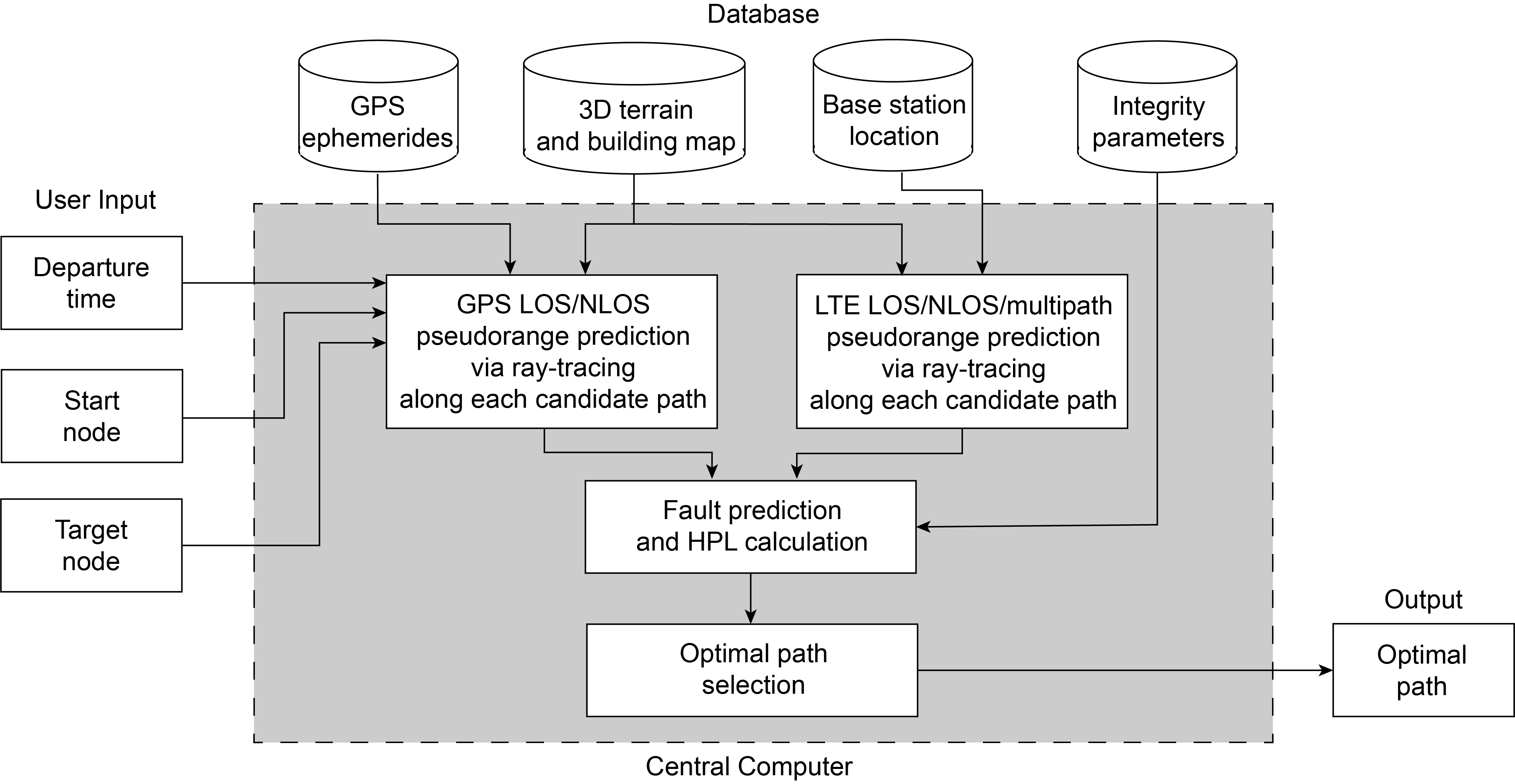}
\caption{Block diagram of the proposed path planning method. The vehicle is assumed to have a database of a 3D terrain and building map, GPS ephemerides, LTE base station locations, and integrity parameters. The user inputs are departure time, a start node, and a target node. Given the user inputs, the central computer performs ray-tracing to simulate GPS and LTE pseudoranges, predicts faults, calculates HPLs along each candidate path, and obtains an optimal path.} \label{fig:BlockDiagram}
\end{figure}

Fig. \ref{fig:BlockDiagram} shows the block diagram of the proposed path planning algorithm. The AGV in this paper is assumed to be equipped with GPS and LTE signal receivers, which can generate pseudorange measurements from GPS and LTE signals, respectively. In addition, the AGV is assumed to have a database containing three-dimensional (3D) terrain and building map, GPS ephemerides over time, and the positions of LTE base stations. The AGV desires to reach a target position by tranversing an optimal path, which considers both the travel distance and position reliability. The metrics for navigation reliability in this study include the HPL and the expected occurrence of faulty signals along the candidate paths. Given a desired destination, an optimal path is calculated that minimizes a cost function, which considers HPL and travel distance while satisfying the following constraints: (i) the ratio of nodes with faulty signals to the total nodes be lower than a maximum allowable ratio and (ii) the HPLs along each candidate path  be lower than the horizontal alert limit (HAL). A ray-tracing simulation is conducted to predict pseudorange biases caused by reflections due to buildings in an urban environment. These biases are used when conducting fault prediction and HPL calculation for all nodes along each candidate path. Ray-tracing predicts GPS and LTE pseudorange measurements at each node. Subsequently, fault prediction and HPL calculation is conducted using the predicted GPS and LTE pseudorange measurements.

\vspace{-0.15cm}


\section{GPS AND LTE PSEUDORANGE PREDICTION} \label{Section:GPS and LTE Measurements}

This section presents the prediction of GPS and cellular pseudoranges, including NLOS and multipath errors.

\subsection{AGV-Mounted Receiver States} \label{SubSection:Receiver States}
\vspace{-0.2cm}
In this paper, it is assumed that the AGV-mounted GPS and LTE receivers can measure pseudoranges from $N$ GPS satellites and $M$ LTE base stations, respectively. Before being driven, the AGV computes predicted pseudoranges using ray-tracing at each node along each candidate path. It is assumed that the coordinates of the GPS satellites and LTE base stations are known to the AGV \textit{a priori}. The coordinates of the GPS satellites are known from past data since the GPS orbit is repeated every 12 hours in sidereal time. In addition, the coordinates of the LTE base station can be known by radio mapping or satellite images.

The 3D positions of the $n$-th GPS satellite and $m$-th LTE base station are denoted as $\vb{r}_{{\mathrm GPS}_n} \triangleq [x_{{\mathrm GPS}_n}, y_{{\mathrm GPS}_n}, z_{{\mathrm GPS}_n}]^{\mathsf{T}}$ and $\vb{r}_{{\mathrm LTE}_m} \triangleq [x_{{\mathrm LTE}_m}, y_{{\mathrm LTE}_m}, z_{{\mathrm LTE}_m}]^{\mathsf{T}}$, respectively. The state vector of the receiver is denoted as $\vb{x}_{\mathrm r} \triangleq [\vb{r}_{\mathrm r}^{\mathsf{T}}, c \delta t_{\mathrm r}]^{\mathsf{T}}$, where $\vb{r}_{\mathsf r}\triangleq[x_{\mathrm r}, y_{\mathrm r}, z_{\mathrm r}]^{\mathsf{T}}$ is the receiver's position, $\delta t_{\mathrm r}$ is the receiver's clock bias, and $c$ is the speed of light.

\subsection{GPS Pseudorange Prediction}\label{SubSection:GPS Measurements}
\vspace{-0.2cm}
The $n$-th GPS pseudorange measurement in an urban environment at time-step $k$, denoted $\rho_{{\mathrm GPS}_n}(k)$, is modeled as
\begin{align}
\begin{split}
\rho_{{\mathrm GPS}_n}(k)&=  {\norm{\vb{r}_{\mathrm r}(k) - \vb{r}_{{\mathrm GPS}_n}(k)}} + c \cdot  \left[\delta t_{\mathrm r}(k) - \delta t_{{\mathrm GPS}_n}(k)\right] \\
&\quad \, +c \cdot \delta t_{{\mathrm iono},n}(k) + c \cdot \delta t_{{\mathrm tropo},n}(k) + \rho_{{\mathrm NLOS},n}(k) + \varepsilon_{{\mathrm GPS}_n}(k),
\end{split}
\end{align}
where $c$ is the speed of light; $\delta t_{{\mathrm GPS}_n}$ is the $n$-th satellite's clock bias; $\delta t_{{\mathrm iono},n}$ and $\delta t_{{\mathrm tropo},n}$ are ionospheric and tropospheric delays, respectively; $\rho_{{\mathrm NLOS},n}$ is the NLOS bias owing to reflection; and $\varepsilon_{{\mathrm GPS}_n}$ is the measurement noise, which is modeled as a zero-mean white Gaussian random sequence with variance $\sigma_{{\mathrm GPS}_{n}}^2$. The terms $\delta t_{{\mathrm GPS}_n}$ and $\delta t_{{\mathrm iono},n}$ can be corrected via the navigation messages, and $\delta t_{{\mathrm tropo},n}$ can be corrected using a tropospheric delay model. The corrected pseudorange of the $n$-th satellite is expressed as
\begin{align}
\begin{split}
  \rho'_{{\mathrm GPS}_n}(k)=  {\norm {\vb{r}_{\mathrm r}(k) - \vb{r}_{{\mathrm GPS}_n}(k)}} + c \cdot  \delta t_{\mathrm r}(k) + \rho_{{\mathrm NLOS},n}(k)+\varepsilon_{{\mathrm GPS}_n}(k).
\end{split}
\end{align}
GPS signals received by the receiver can be divided into two categories: LOS and NLOS. In the LOS case, the term $\rho_{{\mathrm NLOS},n}(k)$ is zero. In the NLOS case, the NLOS bias owing to reflection can be modeled as
\begin{equation}
 {\rho_{{\mathrm NLOS},n}(k)=\rho_{{\mathrm reflection},n}}(k) - {\norm {\vb{r}_{\mathrm r}(k) - \vb{r}_{{\mathrm GPS}_n}(k)}},
\end{equation}
where $\rho_{{\mathrm reflection},n}$ is the total travel distance of the NLOS signal, which can be predicted through a ray-tracing simulation.

\subsection{LTE Pseudorange Prediction}\label{SubSection:LTE Measurements}
\vspace{-0.2cm}
The $m$-th LTE pseudorange measurement in an urban environment at time-step $k$, denoted $\rho_{{\mathrm LTE}_m}(k)$, is modeled as
\begin{align}
\begin{split}
\rho_{{\mathrm LTE}_m}(k)&=  {\norm{\vb{r}_{\mathrm r}(k) - \vb{r}_{{\mathrm LTE}_m}}} + c \cdot  \left[\delta t_{\mathrm r}(k) - \delta t_{{\mathrm LTE}_m}(k)\right] + \rho_{{\mathrm MP},m}(k) + \varepsilon_{{\mathrm LTE}_m}(k),
\end{split}
\end{align}
where $\delta t_{{\mathrm LTE}_m}$ is the $m$-th base station's clock bias, $\rho_{{\mathrm MP},m}$ is the bias owing to NLOS and multipath signals, and $\varepsilon_{{\mathrm LTE}_m}$ is the measurement noise, which is modeled as a zero-mean Gaussian random sequence with variance $\sigma_{{\mathrm LTE}_{m}}^2$. In this paper, it is assumed that the LTE clock bias ${\delta t_{{\mathrm LTE}_m}} (1 \le m \le M)$ can be estimated using a first-polynomial approximation. This is accomplished using a method discussed in \cite{Maaref19:Measurement}.

In general, since the elevation angle of the LTE base station is lower than that of the GPS satellite, the LTE signal can be affected more significantly by a building than the GPS signal, resulting in a greater number of multipath signals. Accordingly, in the case of LTE pseudorange prediction, bias owing to the NLOS and multipath signals is considered. In this paper, it is assumed that LTE pseudoranges are measured by the code phase discriminator developed in \cite{Shamaei17:LTE}. The bias owing to multipath can be calculated using the complex channel impulse response (CIR) at time-step $k$, which is expressed as \cite{Shamaei17:LTE}:
\begin{equation} \label{eq:CIR}
 h(k,\tau)=\sum^{L-1}_{l=0}{\alpha(k,l)\delta(\tau-\tau(k,l))},
\end{equation}
where $L$ is the number of multipath components; $\alpha(k,l)$ and $\tau(k,l)$ are the relative amplitude and delay components of the $l$-th path with respect to the first path, respectively; and $\delta(\cdot)$ is the Dirac delta function. In addition, in (\ref{eq:CIR}), the multipath fading channel is assumed to stay constant over the duration of a symbol \cite{Shamaei17:LTE}.

The complex impulse response can be simulated by a ray-tracing software. Using the complex impulse response, the multipath and NLOS bias, $\rho_{{\mathrm MP},m}(k)$, can be calculated as
\begin{equation}
\rho_{{\mathrm MP},m}(k)=c \cdot \tau(k,0)-{\norm{\vb{r}_{\mathrm r}(k) - \vb{r}_{{\mathrm LTE}_m}}}+\chi_{1}(k)+\chi_{2}(k),
\end{equation}
where $\tau(k,0)$ is the time delay of the first path. When the LOS signal is received, $c \cdot \tau(k,0)$ equals the geographical distance between the receiver and base station, ideally. Further, the multipath channel impacts on the discriminator function, $\chi_{1}(k)$ and $\chi_{2}(k)$, can be calculated as \cite{Yang00:Timing, Shamaei17:LTE}:

\begin{equation}
\chi_{1}(k)=C\left|{\sum^{B-1}_{b=0}\sum^{L-1}_{l=1}{\alpha(k,l)}e^{-2j\pi(b/B)(\tau(k,l)/T_{\mathrm s}+ e -\xi)}}\right|^2
- C\left|{\sum^{B-1}_{b=0}\sum^{L-1}_{l=1}{\alpha(k,l)}e^{-2j\pi(b/B)(\tau(k,l)/T_{\mathrm s}+ e +\xi)}}\right|^2,
\end{equation}

\begin{align}
\begin{split}
\chi_{2}(k)&=2C\Re{\Bigg\{ \left[\sum^{B-1}_{b=0}e^{-j2\pi(b/B)(e- \xi)}\right]\cdot
\left[\sum^{B-1}_{b'=0}\sum^{L-1}_{l=1}{\alpha^{\ast}(k,l)}e^{-2j\pi(b'/B)(\tau(k,l)/T_{\mathrm s}+e-\xi)}\right] \Bigg\} } \\
& - 2C\Re{\Bigg\{ \left[\sum^{B-1}_{b=0}e^{-j2\pi(b/B)(e+\xi)}\right]\cdot \left[\sum^{B-1}_{b'=0}\sum^{L-1}_{l=1}{\alpha^{\ast}(k,l)}e^{-2j\pi(b'/B)(\tau(k,l)/T_{\mathrm s}+e+\xi)}\right] \Bigg\}},
\end{split}
\end{align}

where $\Re\{\cdot\}$ denotes the real part, and $T_{\mathrm s}$ is the sampling interval, which is defined as $T_{\mathrm s} \triangleq T_{\mathrm symb}/N_{\mathrm c}$, where $N_{\mathrm c}$ is the total number of subcarriers (2048 when the bandwidth is 20 MHz), $T_{\mathrm symb} \triangleq 1/\Delta f$, and $\Delta f=15$-kHz spacing is assigned between subcarriers in orthogonal frequency division multiplexing (OFDM). Furthermore, $B$ is the number of subcarriers in the cell-specific reference signal (CRS), and $B=200$ in this paper because the bandwidth of the LTE signal is assumed to be 20 MHz. $C$ is the signal power owing to the antenna gain and implementation loss \cite{Shamaei17:LTE}, $\xi$ is the time shift in the tracking loop ($\xi=0.5$ is chosen in this paper), and $e$ is the normalized symbol error ($e$ is set to zero in this paper while assuming perfect tracking).

\section{FAULT PREDICTION AND HPL CALCULATION}\label{Section:HPL Calculation}
A fault detection and HPL calculation method based on the multiple hypothesis solution separation (MHSS) algorithm\cite{Blanch12:Advanced, Kropp14:Optimized} is used. The MHSS algorithm, unlike certain RAIM algorithms, can consider multiple faults. Because the probability of having large biases in pseudoranges for both LTE and GPS signals in a high-building environment could be high, multiple faults need to be considered. Therefore, using the MHSS-based fault detection and HPL calculation method is appropriate for AGV navigation with GPS and LTE signals in urban areas.

The basic idea of the solution separation technique is to create multiple hypothesis tests that account for the difference between the all-in-view position solution and the fault-tolerant position solution\cite{Kropp14:Optimized}. The fault detection and HPL calculation methods in this section adapt the methods in  \cite{Blanch12:Advanced, Kropp14:Optimized}.

\subsection{Fault-Free and Fault-Tolerant Position Solutions} \label{SubSection:WLS Estimator}
\vspace{-0.2cm}
The fault-free position solution can be obtained from all-in-view satellites from weighted least squares (WLS) as
\begin{equation}
  \Delta \vb{x}^0_{\mathrm r}= \mb{S}^{0} \Delta \vb{\rho},
\end{equation}
where $\mb{S}^{0} \triangleq { \left(    \trp{\mb{G}} {\mb{W}}  \mb{G}   \right) }^{-1} \trp{\mb{G}} {\mb{W}}$; $\vb{\rho}=[\rho'_{{\mathrm GPS}_1}, \ldots, \rho'_{{\mathrm GPS}_N}, \rho_{{\mathrm LTE}_1}, \ldots, \rho_{{\mathrm LTE}_M}]$; and $\Delta \vb{x}^0_{\mathrm r} \triangleq \vb{x}^0_{\mathrm r} -\hat{\vb{x}}^0_{\mathrm r}$ is the difference between the receiver's state vector $\vb{x}^0_{\mathrm r}$ and its estimate $\hat{\vb{x}}^0_{\mathrm r}$, and $\Delta \vb{\rho} \triangleq \vb{\rho} - \vb{\hat{\rho}}$ is the difference between the pseudorange vector $\vb{\rho}$ and its estimate $\hat{\vb{\rho}}$. The geometry matrix is defined as $\mb{G}\triangleq\trp{[\trp{\mb{G}_{\mathrm GPS}}, \trp{\mb{G}_{\mathrm LTE}}]}$, where the $i$-th row of the $\mb{G}_{\mathrm GPS}$ and $\mb{G}_{\mathrm LTE}$ is defined as
\begin{equation} \label{eq:WLS}
  \mb{G}_i= [-\cos{El_i}\sin{Az_i}\;  -\cos{El_i}\cos{Az_i}\;  -\sin{El_i}\; 1],
\end{equation}
where $El_i$ and $Az_i$ refer to the elevation angle and azimuth angle, respectively, to the $i$-th GPS satellite or LTE base station. Furthermore, the weighting matrix $\mb{W}$ is defined as
\begin{equation}
\mb{W} = {\mathrm diag}\left[\sigma^2_{{\mathrm GPS}_1},\ldots,\sigma^2_{{\mathrm GPS}_N},
\sigma^2_{{\mathrm LTE}_1},\ldots,\sigma^2_{{\mathrm LTE}_M}\right]^{-1},
\end{equation}
where ${\mathrm diag}(\cdot)$ denotes a diagonal matrix.

In a similar way, the fault-tolerant position solution is calculated by the WLS estimator. The fault-tolerant position solution of the $i$-th fault mode is obtained from
\begin{equation}
  \Delta \vb{x}^i_{\mathrm r}= \mb{S}^{i} \Delta \vb{\rho},
\end{equation}
where $\mb{S}^{i} \triangleq { \left(    \trp{\mb{G}} \mb{{R}}_{i} {\mb{W}}  \mb{G}   \right) }^{-1} \trp{\mb{G}} \mb{{R}}_{i} {\mb{W}}$, and $\mb{{R}}_{i}$ is an $(N+M) \times (N+M)$ identity matrix in which the zero terms correspond to the faulty signals of the $i$-th fault mode.

\subsection{Fault Prediction} \label{SubSection:Fault Prediction}
\vspace{-0.2cm}
Fault detection in the MHSS algorithm is conducted using the difference between the all-in-view solution and the $i$-th fault-tolerant solution corresponding to the east (${q}=1$) and north (${q}=2$) axes. This is denoted as ${\Delta r_{i,q}}\triangleq \left|\hat{r}^i_{{\mathrm r},q}-\hat{r}^0_{{\mathrm r},q}\right|$. The faulty signals can be detected by an exclusion test\cite{Blanch12:Advanced}:
\begin{align}
\begin{split} \label{eq:Fault Detection}
&\theta_i = 1\quad \mathrm{if} \;\; \Delta \mathnormal{r_{i,q} \leq {D}_{q}^{i}} \;\; \mathrm{for\;all} \; \mathnormal{q} \\
&\theta_i = 0\quad \mathrm{otherwise}
\end{split}
\end{align}
where ${D}_{q}^{i}$ is the detection threshold for the difference between the $i$-th subset position solution and the all-in-view position solution, which is given in \cite{Blanch12:Advanced, Kropp14:Optimized}. A value of $\theta_i=1$ in (\ref{eq:Fault Detection}) indicates a risk that the wrong satellite has been excluded\cite{Blanch12:Advanced}. On the other hand, a value of $\theta_i=0$ means that exclusion of faulty signals corresponding to the $i$-th fault mode should be attempted.

Since this paper aims to predict faulty signals before the vehicle is driven, the conventional fault detection and exclusion (FDE) with live signals, which is suitable for real-time operations, is not conducted. For the purpose of path planning, we focus on predicting faults before driving based on the predicted GPS and LTE signals from the ray-tracing simulation.

\subsection{HPL Calculation} \label{SubSection:HPL Calculation}
\vspace{-0.2cm}
The protection levels corresponding to the all-in-view fault-free position solution for the east (${q}=1$) and north (${q}=2$) axes, $({\mathrm PL}^{0})_{q}$, are obtained from
\begin{equation}
({\mathrm PL}^{0})_{q}={\mathrm K}^{0}_{{\mathrm MD},q}\cdot\sigma_{q}^{0}+
\sum^{N+M}_{j=1}|(\mb{S}^{0})_{q,j}|\cdot{b}_{{\mathrm int},j},
\end{equation}
where $\sigma_{q}^{0}\triangleq {\sqrt{ \left(    \trp{\mb{G}} {\mb{W}}  \mb{G}   \right) ^{-1}_{q,q}}}$, and $\mb{\mathbf{(\cdot)}}_{i,j}$ denotes the element of the $i$-th row and $j$-th column of a matrix. Furthermore, ${b}_{{\mathrm int},j}$ is the maximum nominal bias of the range measurement for the $j$-th satellite or base station when the integrity is considered, and ${\mathrm K}^{0}_{{\mathrm MD},q}$ is the scale factor that satisfies the missed detection probability. This is calculated as
\begin{equation}
{{\mathrm K}^{0}_{{\mathrm MD},1}={ \mathrm K}^{0}_{{\mathrm MD},2}=-{Q}^{-1}
{\left(\frac{{P}_{{\mathrm HMI}}}{{4}\cdot(N_{\mathrm maxsub}+1)}\right)}},
\end{equation}
where $P_{\mathrm HMI}$ is the probability of hazardously misleading information that satisfies the integrity requirement, and $N_{\mathrm maxsub}$ is the total number of subset geometries. Further, ${Q}$ denotes the tail probability function of the standard normal distribution.

The protection level corresponding to the fault-tolerant solution of the $i$-th fault mode in the $q$ direction, $({\mathrm PL}^{i})_{q}$, is obtained from
\begin{equation}
{({\mathrm PL}^{i})_{q}={\mathrm K}^{i}_{{\mathrm MD},q}\cdot\sigma_{q}^{i}+
\sum^{N+M}_{j=1}|\mb{S}^{i}_{q,j}|\cdot{b}_{{\mathrm int},j}}
+{D}_{q}^{i}, \label{eq:Dqi}
\end{equation}
where $\sigma_{q}^{i}\triangleq {\sqrt{ \left(    \trp{\mb{G}} \mb{R}_i {\mb{W}}  \mb{G}   \right) ^{-1}_{q,q}}}$. The scale factor that satisfies the missed detection probability for the $i$-th fault mode in the $q$ direction, ${\mathrm K}^{i}_{{\mathrm MD},q}$, that considers the prior probability of the $i$-th fault mode occurring, ${P}^{i}_{\mathrm f}$, is calculated as
\begin{equation}
{{\mathrm K}^{i}_{{\mathrm MD},1}={\mathrm K}^{i}_{{\mathrm MD},2}=-{Q}^{-1}
{\left(\frac{{P}_{{\mathrm HMI}}}{{4}\cdot{{P}^{i}_{\mathrm f}}\cdot(N_{\mathrm maxsub}+1)}\right)}}.
\end{equation}
The final PL solution for each axis is the maximum value of the PLs calculated from both the fault-free and fault-tolerant modes according to
\begin{equation}
{\mathrm PL}_{q}=\max \left\{ ({\mathrm PL}^{0})_{q},\max_{i}(({\mathrm PL}^{i})_{q}) \right\}.
\end{equation}
The HPL at each node along each candidate path, which is used as the path planning metric in this paper, is defined as follows
\begin{equation}
{\mathrm HPL}=\sqrt{{\mathrm PL}_{1}^{2}+{\mathrm PL}_{2}^2}.
\end{equation}

\section{OPTIMAL PATH SELECTION}\label{Section:Optimal Path Selection}
This section deals with a set of objectives and a formulation of the proposed path planning problem. The optimization problem in this paper accounts for the ratio of nodes with faulty signals to the total nodes, HPLs, and travel distance.

\begin{list}{}{\leftmargin=1em \itemindent=0em}
    \item[1.] \textit{The ratio of nodes with faulty signals to the total nodes}: The existence of faulty signals can cause an erroneous and intolerable position solution. In order to increase the accuracy and reliability of navigation, it is recommended to choose a path that avoids faulty signals. Therefore, paths in which the ratio of nodes with faulty signals to the total nodes exceed the maximum allowable threshold that is specified are excluded.
    \item[2.] \textit{HPL}: The HPLs can represent the reliability of navigation solutions along the given path. This reliability information is useful because the quality of navigation solutions between two different paths can be significantly different although their travel distances may be similar in urban areas owing to the multipath and NLOS environment. For AGV navigation, a path with higher-quality navigation solutions is desirable, so the HPLs are considered in the proposed path planning algorithm. In addition, by excluding paths in which HPL exceeds HAL, the reliability of the navigation solution is maintained at a quality above a certain threshold.
    \item[3.] \textit{Travel distance}: Travel distance is considered as an objective. This avoids the selection of a path that is too long.
\end{list}

Given candidate paths, the optimal path is determined by solving the following optimization problem.
\begin{align}
\begin{split} \label{eqn:optn}
&\underset{\pi \in \mathcal{P}}{\text{minimize}} \quad \sum_{p_k \in \pi} {\mathrm dist}(p_k)\cdot {\mathrm HPL}(p_k,t)
 \\
&\text{subject to} \quad \frac{\sum_{k=1}^{n_{nodes}}{\mathrm fault}(p_k,t)}{n_{\mathrm nodes}} \leq {\mathrm fault}_{\mathrm max} \;\; \text{and} \;\; {\mathrm HPL}(p_k,t) \leq {\mathrm HAL},
\end{split}
\end{align}
where ${\mathrm dist}(p_k)$ is the length between nodes $p_k$ and $p_{k-1}$; ${\mathrm HPL}(p_k,t)$ is the predicted HPL at node $p_k$ at time $t$; ${\mathrm fault}(p_k,t)$ indicates the existence of a faulty signal at node $p_k$ at time $t$ (0 when a faulty signal does not exist and 1 when a faulty signal exists); $n_{\mathrm nodes}$ is the total number of nodes along a given path; ${\mathrm fault}_{\mathrm max}$ is the maximum allowable ratio of the nodes with faulty signals to the total nodes; and ${\mathrm HAL}$ is the maximum allowable horizontal position error under the given integrity requirement. Further, a path from the start to a target node is denoted by $\pi \in \mathcal{P}$, where $\mathcal{P}$ is the set of all possible paths. The path $\pi$ is represented by a sequence of node indices between the start node index $p_{\mathrm s}$ and the target index $p_{\mathrm t}$, namely, $\pi = \{p_{\mathrm s}, p_1, p_2, \ldots, p_{\mathrm t}\}$.


\section{SIMULATION RESULTS}\label{Section:Simulation Results}
This section presents results from the simulation study for the proposed path planning strategy. The integrity parameters were according to Table \ref{tab:settings}. Ray-tracing was conducted using the commercial software package Wireless InSite \cite{Remcom20:www}, and the commercial 3D terrain and building map from 3dbuildings\cite{3Dbuildings20:www}. When ray-tracing the GPS signals, diffraction and penetration are not considered, and only a single reflection is considered for computational efficiency. Thus, only the direct path and a single reflected path are utilized. For LTE signals, multiple reflections are considered because multipath is more pronounced than the case of GPS, but signals with path losses greater than the threshold are excluded (the path loss threshold in this paper is set to $-130$ dB). It is assumed that all walls and floors are made of concrete in the simulated ray propagation environment.

\begin{table} [H]
\centering
\caption{Integrity Parameters} \label{tab:settings}
\vspace{-5mm}
\begin{center}
{\renewcommand{\arraystretch}{1.4}
 \begin{tabular}[c]{>{\centering\arraybackslash}m{2.5cm} >{\centering\arraybackslash}m{8cm}>{\centering\arraybackslash}m{3.5cm} }
 \Xhline{2\arrayrulewidth}
\textbf{Parameter} &\textbf{Definition}&\textbf{Value} \\
 \hline
 ${b}_{{\mathrm int},j}$ & Maximum nominal bias on range measurement for the $j$-th satellite or base station &$0.5$ m \\
 $P_{\mathrm HMI}$ & Probability of hazardously misleading information & $0.01$ \\
 ${P}^{i}_{\mathrm f}$ & Prior probability of the $i$-th fault mode occurring & $2\times10^{-4}$ \\
 $N_{\mathrm maxsub}$ & Total number of subset geometries & $\sum^{3}_{k=1}{N+M \choose k}$ \newline (i.e., up to three faults are considered)  \\
 ${\mathrm fault}_{\mathrm max}$ & Maximum allowable ratio of the nodes with faulty signals to the total nodes & $0.15$  \\
 ${\mathrm HAL}$ & Horizontal alert limit & $40$ m  \\
 \Xhline{2\arrayrulewidth}
\end{tabular}}
\end{center}
\end{table}

\begin{figure}[H]
\centering
\includegraphics[width=0.875\linewidth]{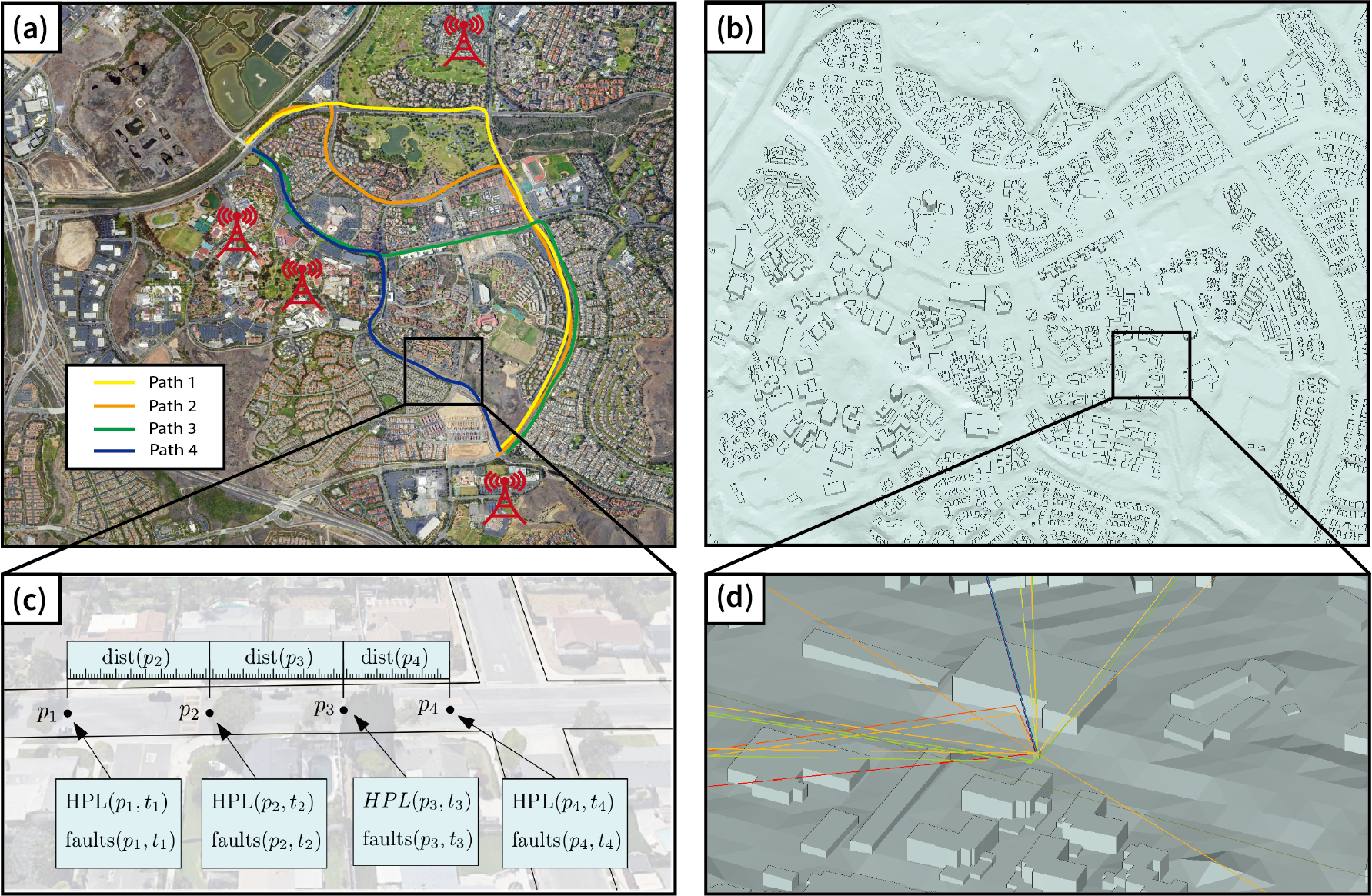}
\caption{(a) Four candidate paths between the start and target node. It is assumed that LTE signals are received from four base stations shown on the map (map image is exported from Google Earth\cite{GoogleEarth20:www}), (b) 3D terrain and building map of the simulation area (University of California, Irvine), (c) Four example nodes with ${\mathrm HPL}(p_k,t)$, ${\mathrm faults}(p_k,t)$, and ${\mathrm dist}(p_k)$, (d) ray-tracing example at a single node.} \label{fig:map}
\end{figure}

Fig. \ref{fig:map}(a) shows four candidate paths and the locations of the LTE base stations. The locations of the LTE base stations were obtained from Cellmapper \cite{Cellmapper20:www}. Also, Fig. \ref{fig:map}(b) depicts the 3D terrain and building map of the simulation area and Fig. \ref{fig:map}(c) presents a ray-tracing example. The GPS ephemerides corresponding to the driving start time of 4:23 pm, January 22, 2020 in Coordinated Universal Time (UTC) were used, and the UGV's velocity was set to 40 km/h. The fault predictions and HPL calculations were conducted at all the nodes along the four candidate paths.

\begin{figure} [H]
\centering
\includegraphics[width=\linewidth]{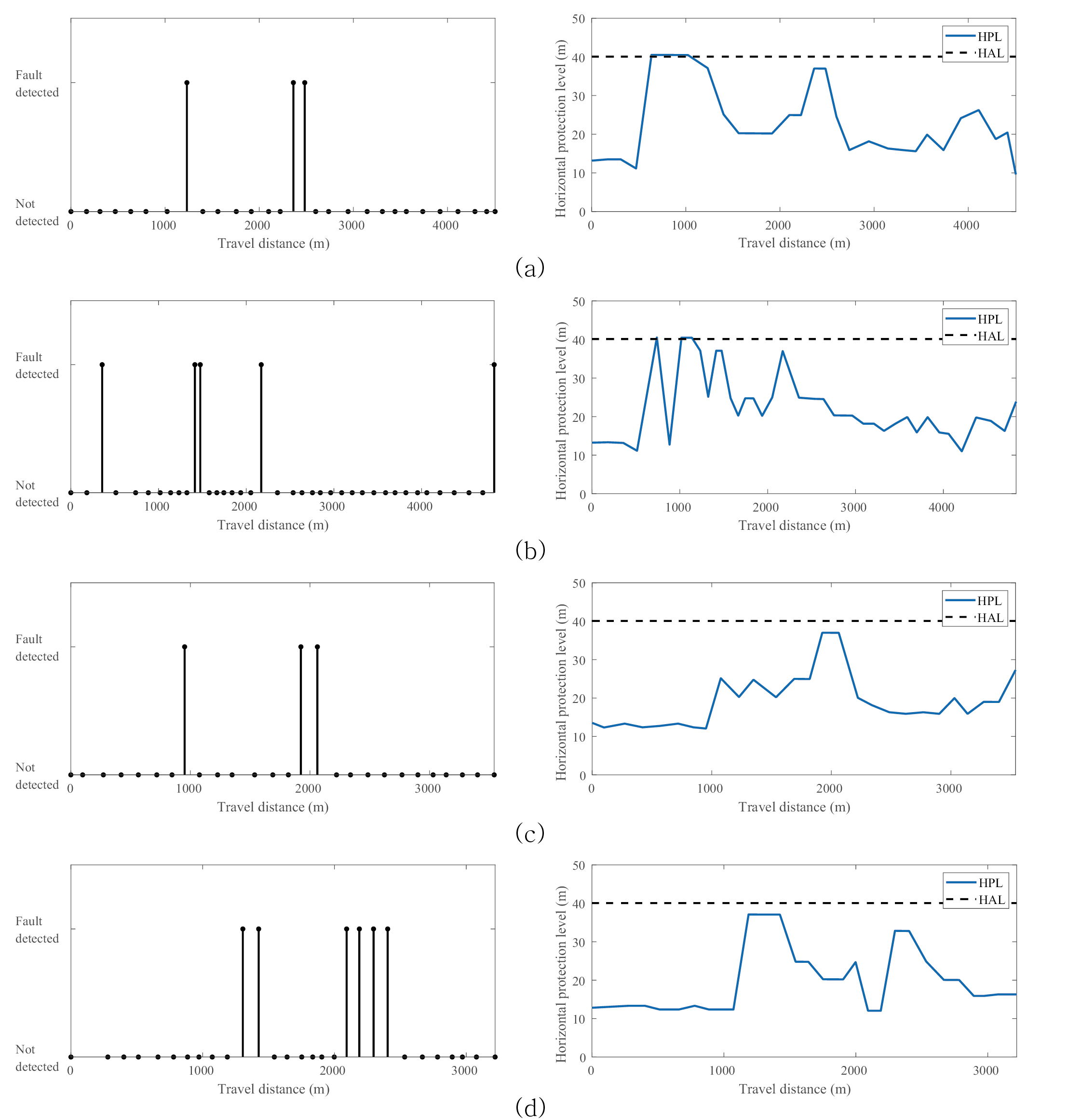}
\caption{Existence of a faulty signal (left) and HPLs (right) along a candidate path: (a) path 1, (b) path 2, (c) path 3 (optimal), (d) path 4 (shortest).} \label{fig:graphs}
\end{figure}

Fig. \ref{fig:graphs} shows the fault prediction and HPL calculation results for each candidate path.  Table \ref{tab:simResults} shows the travel distance, average HPL, maximum HPL, the ratio of nodes with faulty signals, and value of the cost function defined in (\ref{eqn:optn}) of each path. The optimal path determined by the proposed path planning strategy is Path 3, while Path 4 was determined by Google Maps as the shortest path \cite{GoogleMaps20:www}.

\begin{table} [H]
\centering
\caption{Travel distance, average HPL, maximum HPL, ratio of nodes with faulty signals, and value of the cost function of the four candidate paths}\label{tab:simResults}
\vspace{-5mm}
\begin{center}
{\renewcommand{\arraystretch}{1.4}
 \begin{tabular}[c]{>{\centering\arraybackslash}m{2.8cm} >{\centering\arraybackslash}m{2.6cm}>{\centering\arraybackslash}m{2.4cm}>{\centering\arraybackslash}m{2.6cm}
 >{\centering\arraybackslash}m{2.6cm}>{\centering\arraybackslash}m{2.4cm}}
 \Xhline{2\arrayrulewidth}
 \multicolumn{1}{>{\centering\arraybackslash}m{2.8cm}}{\textbf{Path}}
    & \multicolumn{1}{>{\centering\arraybackslash}m{2.6cm}}{\textbf{Travel distance [m]}}
    & \multicolumn{1}{>{\centering\arraybackslash}m{2.4cm}}{\textbf{Average HPL [m]}}
    & \multicolumn{1}{>{\centering\arraybackslash}m{2.6cm}}{\textbf{Maximum HPL [m]}}
    & \multicolumn{1}{>{\centering\arraybackslash}m{2.6cm}}{\textbf{Ratio of nodes with faulty signals [\%]}}
    & \multicolumn{1}{>{\centering\arraybackslash}m{2.4cm}}{\textbf{Cost}}
\\
 \hline
 Path 1  & 4,506 m  & 22.8 m& 40.5 m & 10.3\% & 1,449,761 \\
 Path 2     & 4,828 m  & 22.6 m & 40.5 m & 12.8\% & 1,992,598 \\
 Path 3 (optimal)   & 3,541 m & 19.3 m & 37.0 m & 11.1\% & 994,581  \\
 Path 4 (shortest)  & 3,219 m  & 20.0 m & 37.1 m & 20.7\% & 1,028,836 \\
 \Xhline{2\arrayrulewidth}
\end{tabular}}
\end{center}
\end{table}

\section{CONCLUSION}\label{Section:Conclusions}
This paper presented an integrity-based path planning strategy utilizing GPS and cellular signals for AGV navigation in urban areas. The decision metrics for the integrity-based path planning were (i) the ratio of nodes with faulty signals to the total nodes, (ii) HPL along a candidate path, and (iii) travel time. Ray-tracing using a 3D terrain and building map was performed to predict GPS and LTE pseudoranges in the multipath-rich and NLOS urban environment. With the predicted pseudoranges, fault prediction and HPL calculation were conducted. To enhance the reliability of navigation, candidate paths with the ratio of nodes with faulty signals or HPLs exceeding predefined thresholds were excluded. A simulation study demonstrating the efficacy of the proposed strategy was presented. Among the four candidate paths considered in the simulation study, the path that minimizes the cost function that considered both travel distance and HPLs was selected as the optimal path.


\section*{ACKNOWLEDGMENTS}
The authors would like to thank Mahdi Maaref for insightful discussions. This work was supported in part by the Ministry of Science and ICT (MSIT), Korea, under the High-Potential Individuals Global Training Program (2020-0-01531) supervised by the Institute for Information \& Communications Technology Planning \& Evaluation (IITP). This work was supported in part by the National Science Foundation (NSF) under Grant 1929965.

\bibliographystyle{IEEEtran}
\bibliography{../../../../../../Overmars/Ajax/ajax}

\end{document}